\documentclass[jasr]{elsarticle}
\usepackage{jasr}
\usepackage{amssymb}
\usepackage{latexsym}
\usepackage{graphicx}
\usepackage{natbib}
\usepackage{xcolor}
\usepackage{subfig}
\usepackage[switch]{lineno}
\renewcommand{\arraystretch}{2.5}

\journal{Advances in Space Research}

\begin{document}


\begin{frontmatter}

\title{Data processing of Visible Emission Line Coronagraph Onboard ADITYA–L1}

\author{Muthu Priyal \corref{cor1}}
\cortext[cor1]{Corresponding author$:$ 
  email$:$ muthu.priyal$@$iiap.res.in;} 
  \author{Jagdev Singh, B. Raghavendra Prasad,  Chavali Sumana,  Varun Kumar, Shalabh Mishra, S.N. Venkata, G. Sindhuja, K. Sasikumar Raja, Amit Kumar, Sanal krishnan, Bhavana S. Hegde, D. Utkarsha, Natarajan Venkatasubramanian, Pawankumar Somasundram,  S. Nagabhushana, PU. Kamath, S. Kathiravan, T. Vishnu Mani, Suresh Basavaraju, Rajkumar Chavan, P. Vemareddy, B. Ravindra}
  \author{ S.P. Rajaguru, K. Nagaraju, Wageesh Mishra, Jayant Joshi, Tanmoy Samanta,  Piyali Chatterjee}
  \author{C. Kathiravan,  R. Ramesh}


\address{ Indian Institute of Astrophysics, Koramangala, Bengaluru - 560034}

\begin{abstract}

ADITYA-L1 is India$’$s first dedicated mission to observe the sun and its atmosphere from a halo orbit around L1 point. Visible emission line coronagraph (VELC) is the prime payload on board at Aditya-L1 to observe the sun’s corona. VELC is designed as an internally occulted reflective coronagraph to meet the observational requirements of wide wavelength band and close to the solar limb (1.05 Ro). Images of the solar corona in continuum and spectra in three emission lines 5303\r{AA} [Fe xiv], 7892\r{AA} [Fe xi] and 10747\r{AA} [Fe xiii] obtained with high cadence to be analyzed using software algorithms automatically. A reasonable part of observations will be made in synoptic mode, those, need to be analyzed and results made available for public use. The procedure involves the calibration of instrument and detectors, converting the images into fits format, correcting the images and spectra for the instrumental effects, align the images etc.  Then, develop image processing algorithms to detect the occurrence of energetic events using continuum images. Also derive physical parameters, such as temperature and velocity structure of solar corona using emission line observations. Here, we describe the calibration of detectors and the development of software algorithms to detect the occurrence of CMEs and analyze the spectroscopic data.

\end{abstract}


\end{frontmatter}

\section{Introduction}

A reliable spectroscopic observation with good photometric accuracy and high spectral resolution of the emission corona up to 1.5Ro (Ro - Solar radius) or beyond will help to understand the physical and dynamical nature of solar corona (\citet{singh2006, singh2019}). Most of the spectroscopic or imaging data in visible emission lines have been recorded during the occurrences of total solar eclipses or using ground-based coronagraphs during excellent clear sky conditions. The ground-based coronal observations are insufficient due to limited number of hours per day with very clear coronagraphic sky conditions (\citet{singh2004, singh2011a, ichimoto1999}). The increase in sky brightness because of scattering of sun-light from water vapors and aerosols in the earth’s atmosphere makes it challenging to observe the weak coronal signal. Such a problem does not arise in space, and we can observe 24 hours a day round the year by placing the satellite in a halo orbit around Lagrangian L1 point. 

ADITYA-L1 mission is a space based solar observatory with seven payloads on-board, which is planned to be placed in the first Lagrange point (L1) of the sun-earth system. Visible Emission Line Coronagraph (VELC) (\citet{singh2011} and \citet{prasad2017}) is a space based solar coronagraph, a major payload on-board Aditya-L1 to study the solar corona. The VELC onboard Aditya-L1 is designed to perform imaging of solar corona at 500 nm, simultaneous spectroscopy of solar corona in emission lines centered around 530.3 nm [Fe XIV], 789.2 nm [Fe XI], 1074.7 nm [Fe XIII] and Spectro-polarimetry at 1074.7 nm [Fe XIII]. FOV of the imaging channel is 1.05 - 3.0 Ro and 1.05 – 1.5 Ro for spectroscopy channels. The payload's uniqueness stems from the fact that observations of solar corona closer to the limb (1.05Ro) with a high cadence (\citet{prasad2017} and \citet{kumar2018}) are possible. Further, imaging of the solar corona in continuum will yield the speed of CME$’$s in the plan of sky and spectroscopic emission line observations in the line-of-sight at the same time. Hence, it will be possible to derive the true velocity of CME$’$s and study the acceleration or de-acceleration of CME’s.

Several corrections are required for the solar data obtained from space instruments compared to the data taken by ground based telescope.  There are many space based missions to study the sun and its atmosphere (\citet{brueckner1995} and \citet{wulser2018}).  It is convenient to transmit the data to the ground station from space instruments in binary and compressed format to permit more observations within the same volume of data to be sent through telemetry. Therefore, as a first step of the calibration, the raw data needs to be decompressed and convert to Flexible Image Transport (FITS) format. Then many corrections to the data such as, check file size, dark current, flat fielding of image $/$ spectra, geometrical calibration, wavelength calibration, time-dependent corrections, replacement of spikes $/$ bad pixels and aligning the images are need to be made. The procedure to perform the basic corrections is discussed in earlier paper (\citet{singh2022}). After doing the basic corrections, one needs to convert the observed counts using the digital cameras to absolute numbers such as flux or absolute intensity and correct for the instrument characteristics. The other effects due to scattered light, broadening of line profiles because of instrument, geometrical distortion of spectra because of optics need to be corrected after the basic analysis of data. To apply these corrections, calibration of the instrument in laboratory before the launch and  in space after the launch is needed. Here, we shall discuss the calibration of detectors carried out in the laboratory and development of software to analyse the data to derive physical parameters from the observations made. To examine the performance of the developed software codes we have used the coronal images from C-2 coronagraph onboard SOHO and coronal spectroscopic data obtained with 25cm coronagraph at Norikura Observatory at Japan.  The provision was made to record observations in 4 emission lines, simultaneously, at the 25-cm coronagraph at Norikura observatory, Japan. The 5303\r{AA} [Fe xiv], 10747\r{AA} [Fe xiii] and 10798\r{AA} [Fe xiii] were common to record the spectra. In addition, one of the two, 6374\r{AA} [Fe x] or 7892\r{AA} [Fe xi] spectral lines could be chosen for observations. We have also used the laboratory calibration data to verify the codes and confirm the satisfactory working of the instrument as desired. Here, we discuss the calibration of the detectors, some specific corrections and the verification of the codes using observations with SOHO, Norikura coronagraph and data obtained during total solar eclipses.

\section{Calibration of the Instrument}

The specifications of various optical components of the instrument, performance of mechanical units, responses of detectors need to be examined, individually first and later, the integrated payload are described in \citet{venkata2017, venkata2021}.  There are three CMOS and one IR detectors. The performance of all the three CMOS detector is similar. 

\subsection{ \textbf{Calibration of CMOS detectors in Laboratory before launch}}

The important parameters for a detector to be determined are bias,  dark current variations with exposure time, linearity in response of the detector to light level and exposure time, noise level and signal variation  over pixels with different exposures of same time.

\subsubsection{ \textbf{Bias and dark calibration}}

\begin{figure}[!htbp]
\begin{center}
\includegraphics[width=6.4in, height=3.8in,scale=0.5]{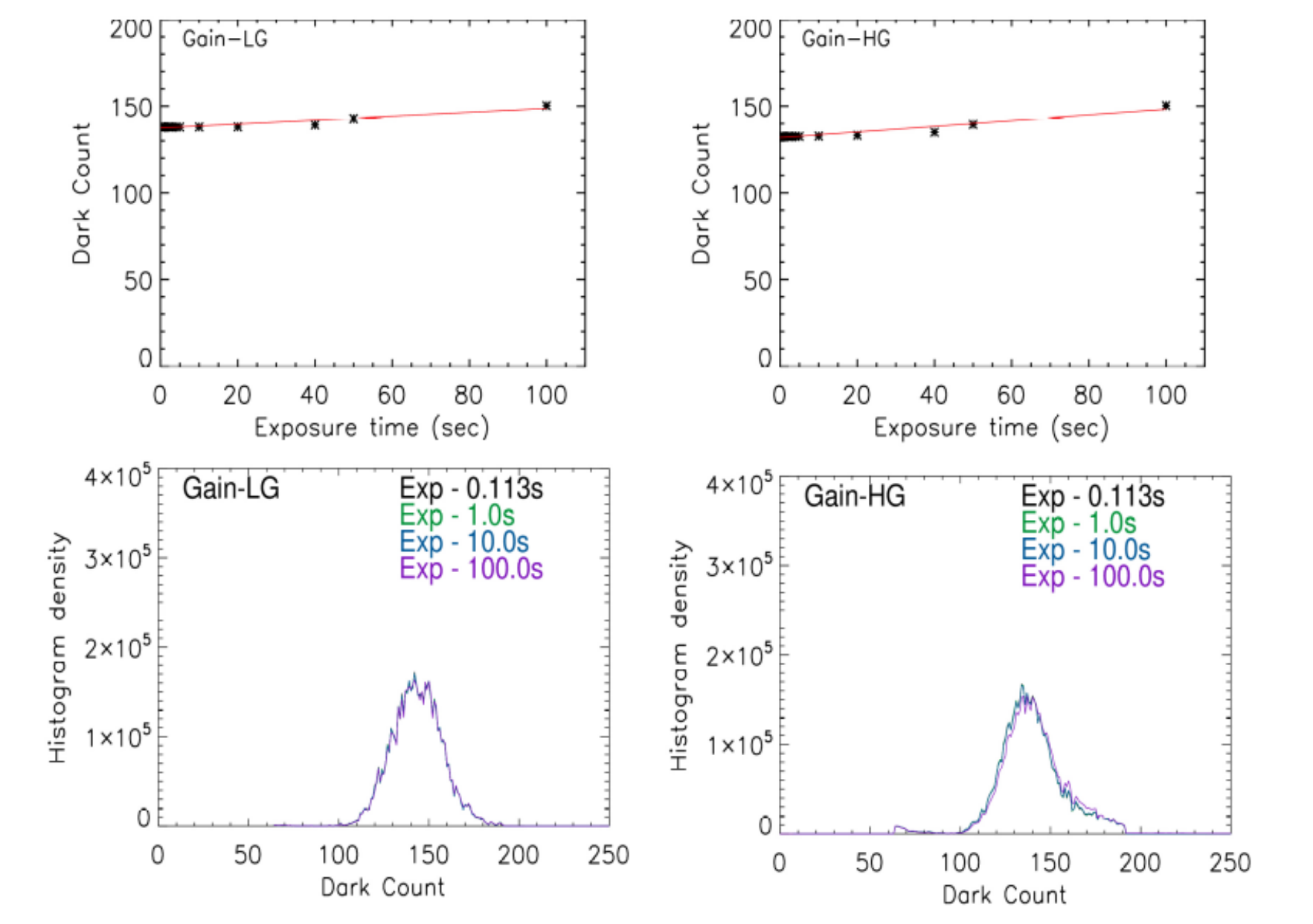}
\caption{The left and right panels in the upper row show the variation of dark current with exposure time for the low(LG) and high(HG) gain of CMOS camera, respectively. The left and right panels in the bottom row show the histogram of the dark current at different exposure times for the low and high gain. }
\label{fig:1}
\end{center}
\end{figure}

The camera electronics does not permit to measure the dark current with zero exposure time known as bias. Therefore, we have taken the images with minimum exposure time of 0.010 s to estimate the bias and its variation with time. We find that mean bias is about 142 and 139 counts in the low gain (LG) and high gain (HG), respectively which does not vary with time. We have measured the bias current at temperatures from -4$^{\circ}$C to -7$^{\circ}$C with an incremental value of 1 deg C. This is the expected operating temperature of CMOS detectors on-board VELC instrument during the mission life and found that these bias values do not change within this temperature range. Two left and right panels in the upper row of Figure \ref{fig:1} show the dark current in counts as a function of exposure time (0.01 to 100 seconds) for the low and high gain, respectively. The dark current plotted in figure is mean over all the pixels in the image. Two panels in the upper panel indicate that dark current increase marginally with exposure time up to 100 seconds. The mean variation in dark current with time indicate that dark build up is at the rate of 0.004  and 0.032 counts $/$ second in the low and high gain, respectively. These variations are insignificant for the projected exposure times. Two panels in the bottom row show the histogram of count for four representative exposure times in the range of 0.113, 1.0, 10.0 and 100 seconds in different colours. The histogram plots indicate that dark current is stable in both the low and high gain for different exposure times. The distribution of dark current agrees very well for the LG and differs by insignificant amount for HG.  The FWHM of the dark current distribution is $\sim$ 30 counts for all the exposure times up to 100 seconds for low and high gains. The small fluctuations in the histograms are due to variation in signal in different rows caused by different amplifiers of CMOS detector. Most of the dark count values range between 100 and 180 for all gains. These variations are mostly due to different gains of amplifiers and photon noise.

\subsubsection{ \textbf{Calibration of CMOS detector with uniform light source in laboratory}}

\begin{figure}[!htbp]
\begin{center}
\includegraphics[width= 6.4in, height=3.2in,scale=0.5]{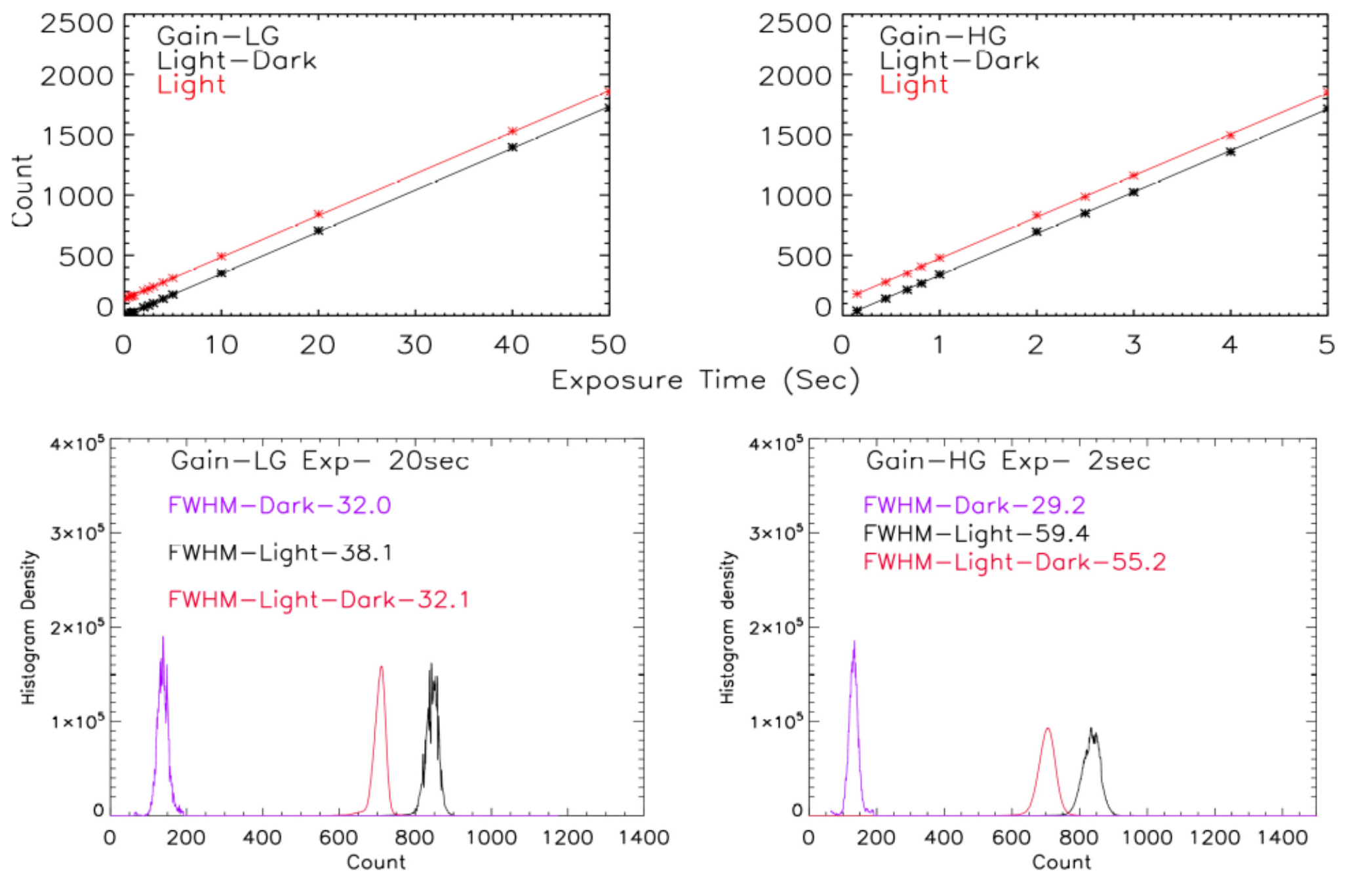}
\caption{Left and right panel in the upper row of the figure show the variation of mean signal in counts for light and (light – dark) with exposure time for the low and high gain for the CMOS camera, respectively. The left and right panels in the bottom row show the histogram of counts for dark (Blue), light (Black) and light-dark (Red) for the low and high gain, respectively.  }
\label{fig:2}
\end{center}
\end{figure}

After taking the dark data we have taken the images with uniform light source at different intensity and exposure times till the near saturation of the detectors. Left and Right side panels in the upper row of Figure \ref{fig:2} show the mean signal over the image (counts) as a function of exposure time for the low and high gains, respectively. Upper curve (red) in both the panels indicate the mean counts with light and lower curve after subtracting averaged dark image of 16 individual images from the light image.  The low and high gain plots indicate that response of the detector is almost linear to the exposure time.  For an exposure time $>$ 100 ms, the increase in signal is linear with time till 90 \% of the saturation value for the detectors. Both the LG (1X and 2X) and HG (10X and 30X) show the similar behaviour. The experiment was repeated with different known intensity levels and found that detectors show linear response in the range of 5 – 90 \% of the full well capacity of the detectors for all the gains.

Left and right panels in the bottom row of Figure \ref{fig:2} show the histograms of the count values for the dark image in blue, image with uniform light source in black and the dark subtracted light image in Red  for LG and HG, respectively. The histograms of dark and light image show some departure from the Gaussian distribution due to fixed pattern noise in the data. But the Gaussian distribution of histogram of (light – dark) image indicates the noise due different response of amplifiers (fixed pattern noise in the data) has been corrected. For the LG image, the FWHM of about 32 counts of the corrected intensity distribution indicates the variation in the signal is well within the photon noise considering average signal count of $\sim$ 850. For the HG data, FWHM of $\sim$ 55 counts for a mean signal of 840 indicates that variations in the signal are almost equal to the photon noise. This is likely due to additional  noise in detector at high gain. All the three CMOS cameras have been calibrated and behave in a similar way. 

\subsection{ \textbf{Calibration of IR detector}}

\begin{figure}[!htbp]
  \centering
  \subfloat[]{\includegraphics[width=5.4in, height=2.5in,scale=0.5,keepaspectratio]{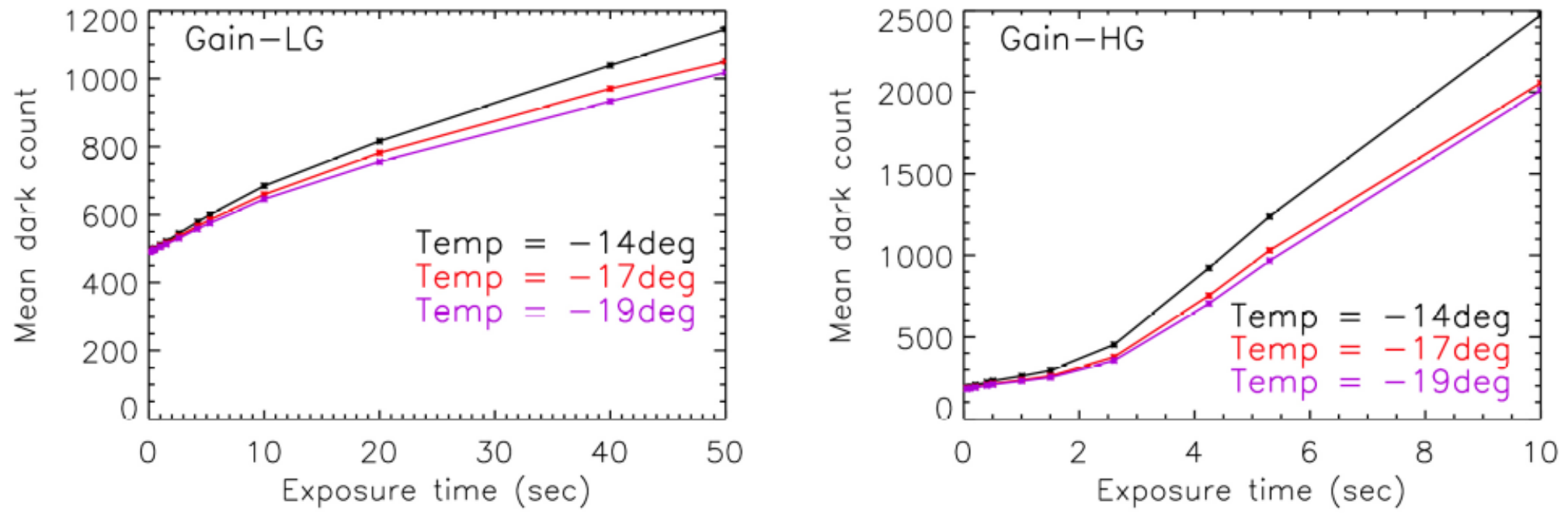} \label{fig:3a}} \\
  \subfloat[]{\includegraphics[width=5.4in, height=2.5in,scale=0.5,keepaspectratio]{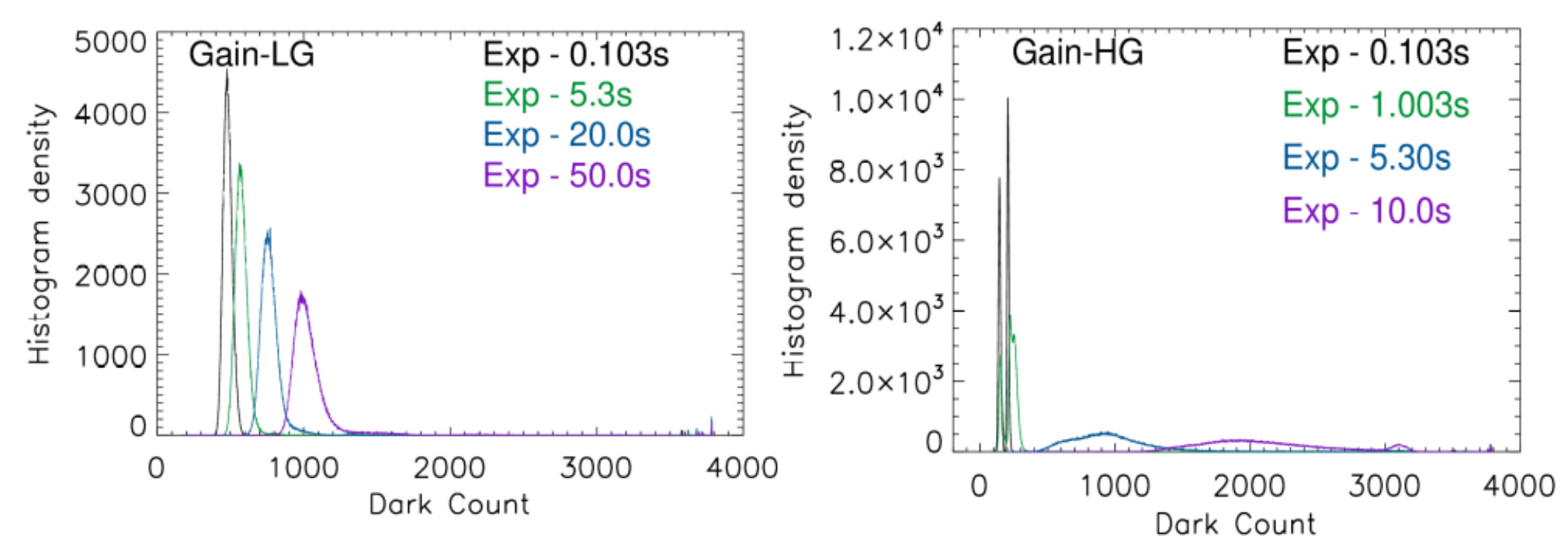} \label{fig:3b}}
  \caption{(a) The left and right panels show the variation of mean dark count over an image with exposure time for the low (LG) high gain (HG), respectively, for the IR camera at temperatures of -14$^{\circ}$C, -17$^{\circ}$C and -19$^{\circ}$C. (b) The top and bottom panels of the figure show the histogram of dark count with exposure times of 103ms, 5s, 20s and 50s for the LG and for HG for exposure time of 103ms, 1s, 5s and 10s for the IR camera at –17$^{\circ}$C.} \label{fig:3}
\end{figure}

\begin{figure}[!htbp]
\begin{center}
\includegraphics[width=5.4in, height=2.5in,scale=0.5]{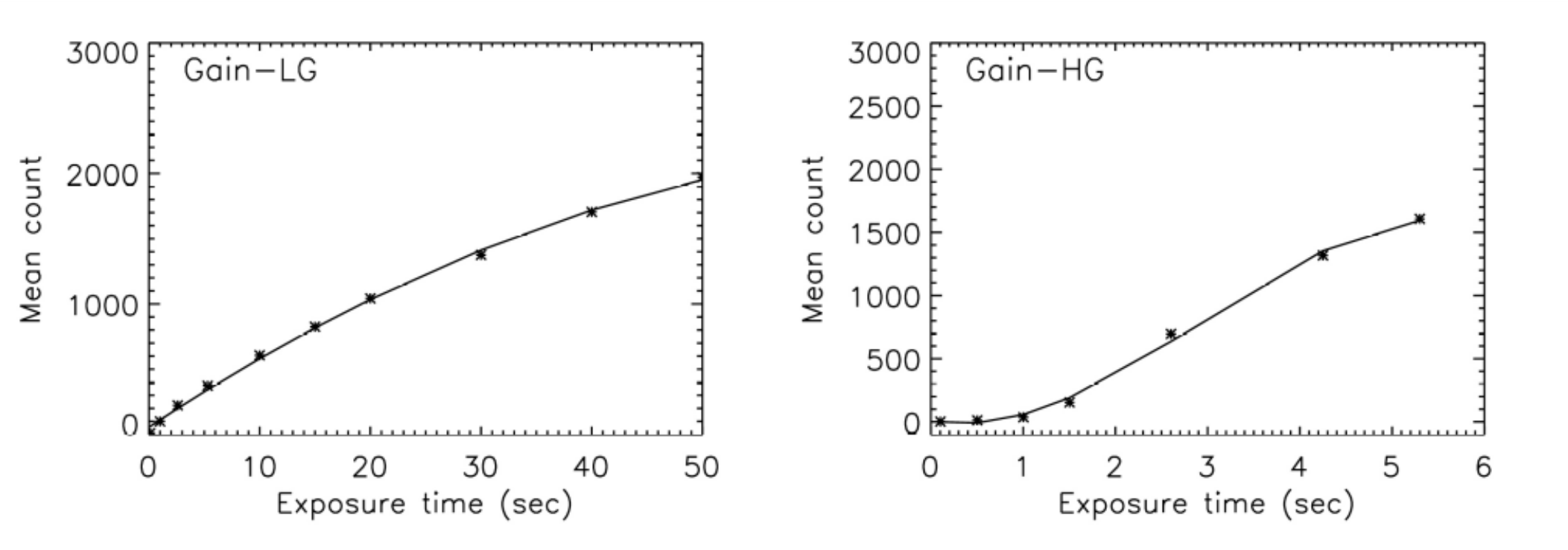}
\caption{Left and right panels of the figure show the mean signal in counts (light – dark) for image with uniform light source as a function of exposure time for IR detector at – 14$^{\circ}$ C for the LG and HG, respectively.}
\label{fig:4}
\end{center}
\end{figure}

First, we study the variation of mean dark current over an image with exposure time for the temperatures from -14$^{\circ}$C to -19$^{\circ}$C, expected range of temperature of the detector onboard, at an interval of one degree. Left and right panels in the upper row of Figure \ref{fig:3}a show the variation of dark count for the LG and HG of the camera, respectively for three temperatures, -14$^{\circ}$C, -17$^{\circ}$C and -19$^{\circ}$C. The figure indicates that dark count is dependent on temperature and increases with increase of temperature of the detector. The difference is less for small exposure times and the difference goes on increasing with the increase in exposure time. Left and right panels of the Figure \ref{fig:3}b show the histogram of the dark current for the LG and HG, respectively, for four exposure times. The exposure times are 103ms (minimum), 5.3s, 20s and 50s (maximum) for LG and 103ms, 1.003s, 5.3s and 10s for HG as the detector gets saturated for exposure $>$ 10s in HG for dark image, itself. The histogram plots for dark current show double peaks. The reason for this is not clear. However, the double peaks disappear in the dark corrected light (light – dark) images. Therefore, the double peak may be due to some fixed pattern noise in the detector.  Further, dark count increases with exposure time significantly unlike the behavior of CMOS detector for which dark count increases negligibly with exposure time up to 100 seconds. The mean dark current is $\sim$ 490 and $\sim$ 190 counts for the  LG and HG, respectively, for an exposure time of 103ms. The histograms of the dark current for LG, for different exposure time show that dark count value increases by $\sim$ 50 \% with an exposure time of 20 seconds as compared to that with 103ms, decreasing the dynamic range significantly. In addition, the increase in width in the distribution of dark count with increasing exposure times indicates significant increase in the dark noise. Thus, the exposure time more than 20 seconds will have impact on dynamic range and photometric accuracy in LG observations. The right side panel of Figure \ref{fig:3}b shows that for exposure time $>$ 2s in HG, the distributions of dark count become very broad and mean value increases at a faster rate. The computed values of dark build up indicate that dark count increase at rate of 12 counts$/$sec in the LG and 220 counts$/$ sec in HG. Hence, it is advisable to keep the exposure time $<$ 5s for HG observations considering the dark build up and noise in the data for larger exposure times. In Figure \ref{fig:4}, we plot the mean signal (Light – dark) in counts for LG (left panel) and HG (right panel) for images with uniform light source, as a function of exposure time.

\section{Observations in continuum channel at 500 nm}

After generating fits files from binary files, the dark current and flat-field corrections, the images (\citet{singh2022}) will be aligned depending on the satellite data about yaw, roll and pitch angles. To begin with all these images will be scanned visually in the form of video to detect the occurrence of CME. 

\subsection{ \textbf{ Detection of CMEs using continuum images}}

We have developed a code to detect the occurrence of CME$’$s, automatically. First, the aligned FITS images of a single day are taken for generating the background image ($I_{bac}$). The procedure involved is to create a minimum image ($I_{min}$) such that each pixel in $I_{min}$ corresponds to the minimum intensity of all the images on that day. The intensity in the outer corona is very less, little more than the dark value. Sometimes signal is lost in the photon noise. Large number images need to be added to increase the signal to noise ratio (SNR) in the outer corona. Sometimes, some pixels, especially in the outer corona in the dark subtracted images show negative values due to photon noise. To avoid this type of noise in an image, the minimum background is taken as zero for those pixels. This generated $I_{min}$ is used to produce the azimuthally averaged background image ($I_{bac}$). To make the background image, the minimum image $I_{min}$ is rotated from 0$^{\circ}$ to 360$^{\circ}$ deg at an increment of 1$^{\circ}$ to get 360 images. Then, we generate the azimuthally averaged image ($I_{bac}$) by averaging over the 360 images such that each pixel in $I_{bac}$ corresponds to the average of all the   intensities over 360 images at that pixel. To detect the occurrence CME (Icme), we subtract the contribution of background from the coronal image to enhance the contrast of the image, using the azimuthally averaged background image ($I_{bac}$) and the following relation$:$

	\begin{equation} \label{eqn}
	I_{cme}= (Image – I_{min}) / I_{bac}
\end{equation}

\par
\begin{figure}[!htbp]
\begin{center}
\includegraphics[width=7.4in, height=2.3in,scale=0.5]{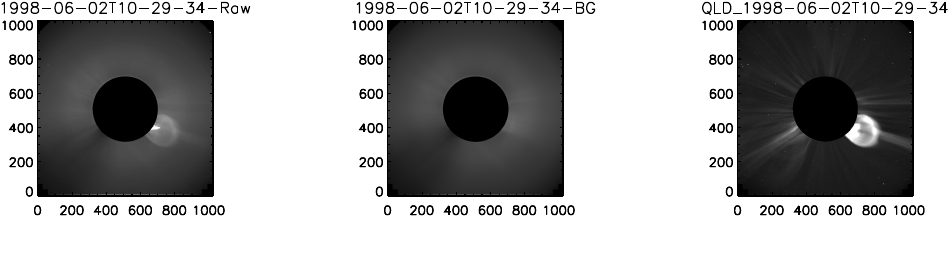}
\caption{The Left and middle panels of the figure shows Level 0.5 image of the LASCO-C2 image obtained on June 2, 1998 at 10:29:34 UT and the background image ($I_{bac}$) generated using all the images of that day. The right side panel shows the CME and coronal streamer structures for this image ($I_{cme}$). The vertical bar shows the relative brightness of various structures. }
\label{fig:5}
\end{center}
\end{figure}

\begin{figure}[!htbp]
\begin{center}
\includegraphics[width=7.4in, height=2.3in,scale=0.5]{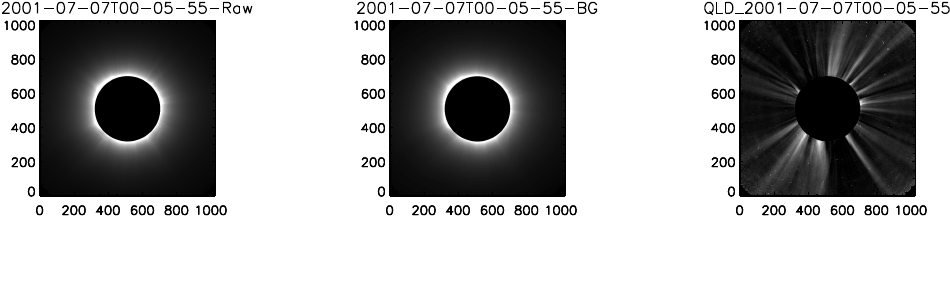}
\caption{The Left and middle panels of the figure shows Level 0.5 image of the LASCO-C2 image obtained on July 7, 2001 at 00:05:55 UT and the background image ($I_{bac}$) generated using all the images of that day. The right side panel shows the coronal streamer structures for this image ($I_{cme}$). The vertical bar shows the relative brightness of coronal structures. }
\label{fig:6}
\end{center}
\end{figure}

 As a case study, we have applied the developed algorithm to broad-band coronal images obtained by LASCO-C2. We have downloaded number of Level 0.5 images from the archive of LASCO-C2, in which north pole is aligned up. We have subtracted the dark offset from the Level-0.5 images and generated the back ground image. Then applied the CME detection code to detect the occurrence
of CME. We found that the code works well and Figure \ref{fig:5}  shows one such example of CME occurrence on June 2, 1998 at 10:29:34 UT. The left and middle panels of the figure show the raw image and the generated background image considering all the images obtained on that day. Figure \ref{fig:6} shows another example of the image obtained on July 7, 2001 at 00:05:55 UT using C2 coronagraph onboard SOHO showing the coronal streamer structures clearly after the analysis. It may be noted that downloaded images does not show the streamer structures. It does indicate the occurrence of faint structure that may be CME, which need to be confirmed. The panel in the right side of the figure shows image ($I_{cme}$) with the detected bright CME$’$s and long lived streamer structures of the solar corona. 

We have tested our algorithm for various events with Lasco-C1 and Lasco-C2 data. Our algorithm works well in detecting the Coronal features like CME$’$s and streamers. We plan to develop the code to determine the speed of CME in the plane of sky. 

\subsection{ \textbf{Merging the LG and HG images}}

\begin{figure}[!htbp]
\begin{center}
\includegraphics[width=6.4in, height=2.7in,scale=0.5]{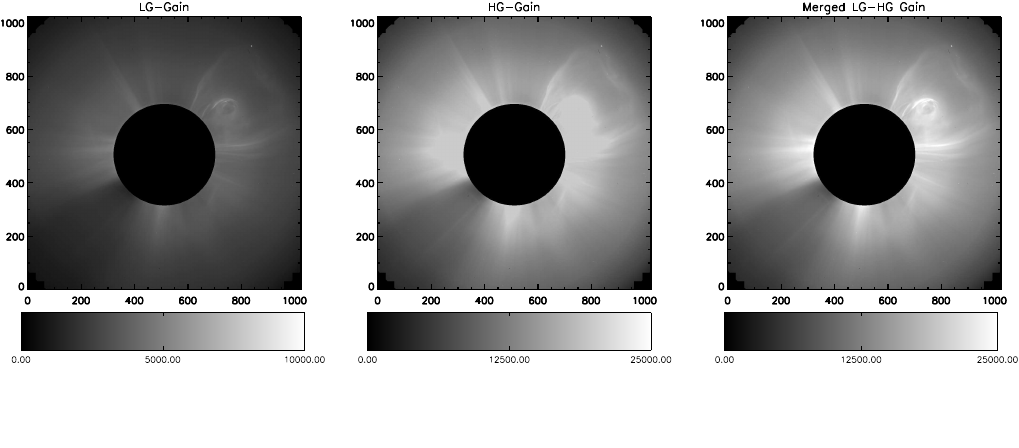}
\caption{ Left panel of the figure shows the LG image, middle panel shows the HG image and right panel of the figure shows merged LG and HG images. }
\label{fig:7}
\end{center}
\end{figure}

Two images of solar corona will be taken, one in LG and other in HG, simultaneously, due to limited dynamic range of CMOS detector and large difference in the intensity in the inner and outer corona. The signal, therefore, is likely to be very less above the dark noise at the outer corona in LG and saturated at the inner corona in HG. Considering the conversion factor of LG and HG, the images will be merged to generate a single image using a software code developed. To develop the code we have used the LASCO-C2 image. We termed the image as a LG as seen in left side panel of Figure \ref{fig:7}. Then generated the HG image by considering the gain factor of 5 (2X and 10X gains in case of VELC) as shown in middle panel of the figure. It may be noted that part of the image is saturated in HG. Considering these images and the gain factor we combined the LG and HG images to generate full image as seen in the right side panel of the figure. It may appear that there is not much difference in the LG and combined image, probably original data is 16-bit format. We expect to see the difference in the 11-bit images taken with VELC in the continuum channel.

\subsection{ \textbf{Equal intensity contour maps of corona}}

\begin{figure}[!htbp]
\begin{center}
\includegraphics[width=5.9in, height=3.2in,scale=0.5]{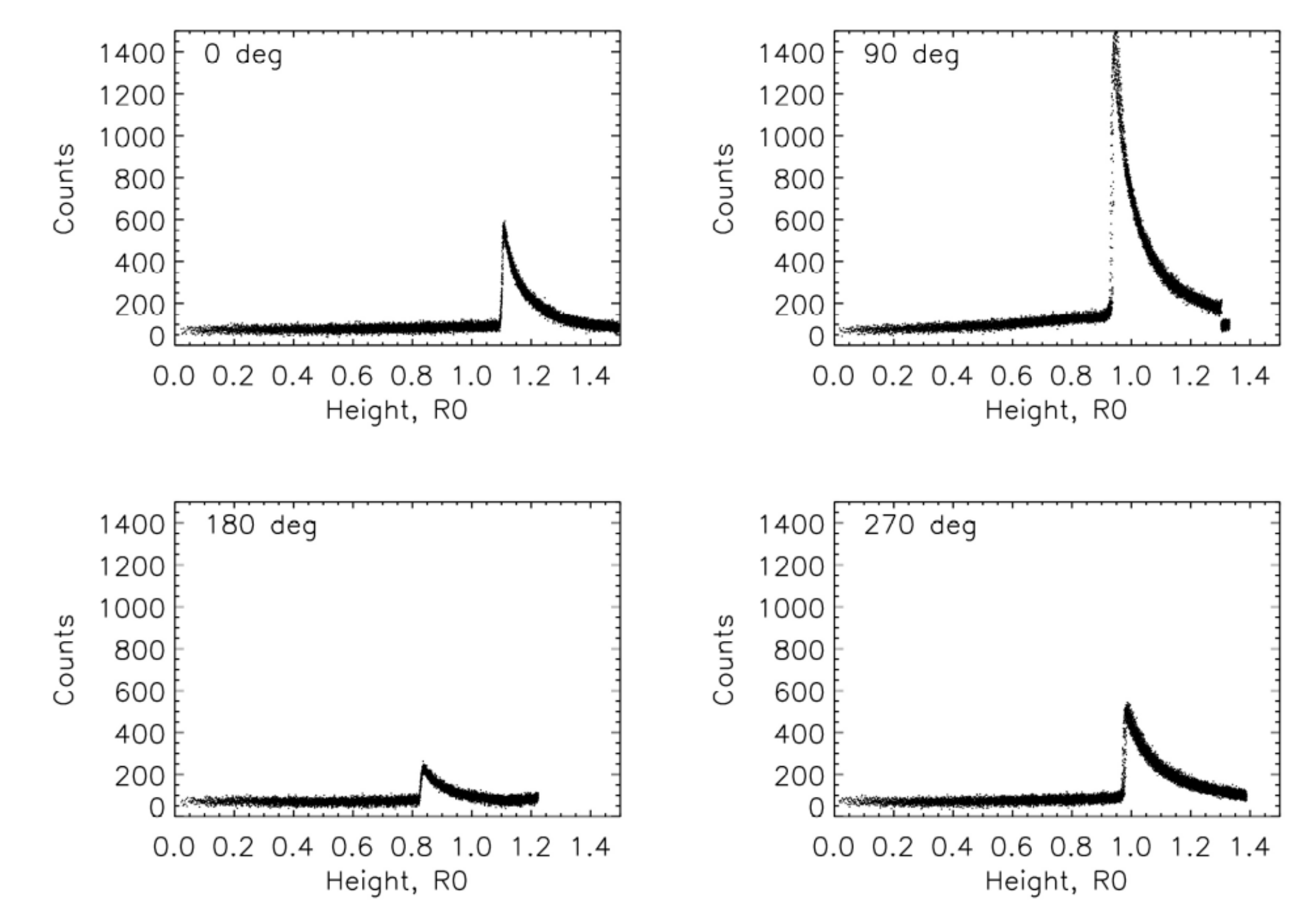}
\caption{ The figure shows the intensity profiles at 0$^{\circ}$, 90$^{\circ}$, 180$^{\circ}$ and 270$^{\circ}$ azimuth angle. }
\label{fig:8}
\end{center}
\end{figure}

\begin{figure}[!htbp]
\begin{center}
\includegraphics[width=3.4in, height=3.2in,scale=0.5]{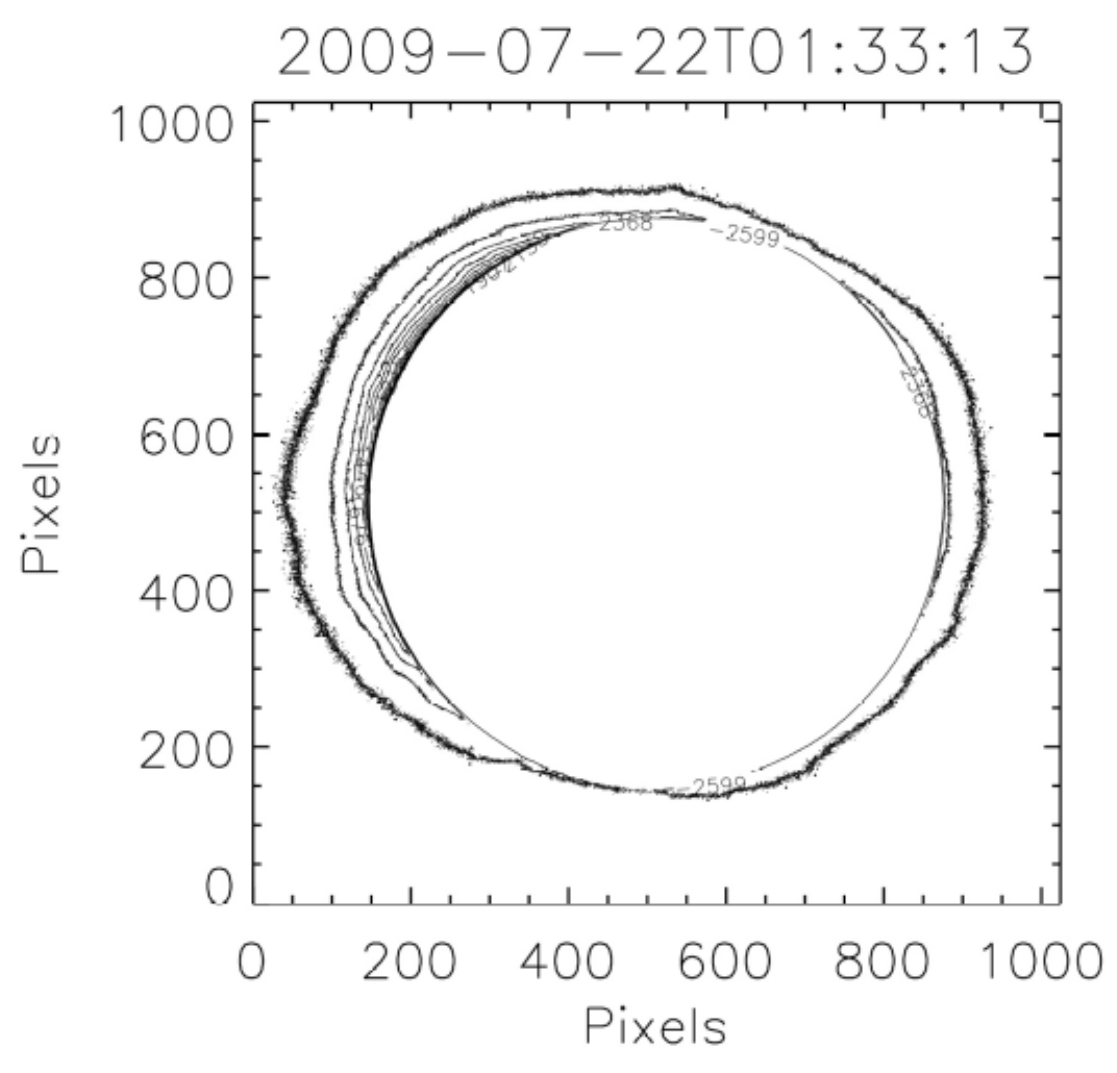}
\caption{The figure shows the equal intensity contour map of coronal images in red emission line during the total solar eclipse of July 22, 2009. }
\label{fig:9}
\end{center}
\end{figure}

We plan to make equal intensity contour maps of solar corona as function of solar radii on daily basis in units using the solar disk observations obtained onboard. We combine the LG and HG images obtained. Average the images over 60 minutes. Using this average image, we make the average intensity profiles as a function of solar radii at an interval of 5$^{\circ}$ in azimuth angle. Normalize these intensity profiles using solar disk data. To develop the code, we used the coronal images in red emission line taken during the total solar eclipse of July 22, 2009. Figure \ref{fig:8} shows the intensity profiles at 0$^{\circ}$, 90$^{\circ}$, 180$^{\circ}$ and 270$^{\circ}$ azimuth angle. Then join the chosen set of equal intensity points at 5$^{\circ}$ interval to generate equal intensity contour map as seen in Figure \ref{fig:9}. It may be noted that Figures \ref{fig:8} and \ref{fig:9} are on relative intensity scale but we plan to make such maps in the scale of solar disk intensity.  The equal intensity contours of the solar corona can be used to study long term solar cycle variations. In addition, variations in the quiet coronal structures can be studied due to occurrence of energetic events.

\section{Analysis of spectroscopic observations}

\begin{figure}[!htbp]
\begin{center}
\includegraphics[width=4.0in, height=3.5in,scale=0.5]{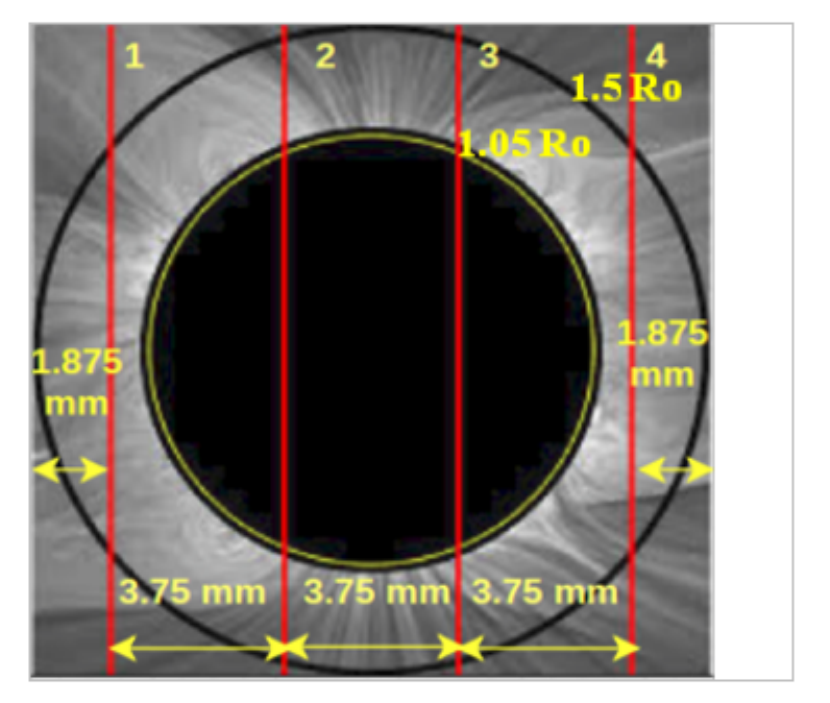}
\caption{The figure  shows the location of 4 slits when LSM is at home position. The slits are 50 micron wide and separated by 3.75 mm. }
\label{fig:10}
\end{center}
\end{figure}

The spectroscopic observations will be made in 3 emission lines, namely 5303\r{AA} [Fe XIV], 7892\r{AA} [Fe XI], 10747\r{AA} [Fe XIII] using multi-slit spectrograph with 4 slits. The figure \ref{fig:10} shows the location of 4 slits when linear scan mechanism (LSM) to move the image on slits in steps, is at home position. The slits are 50 micron wide and separated by 3.75 mm. The image of sun and corona can be moved on the slits using LSM (\citet{singh2022}). One mode of observations is to keep the coronal image fixed on the slits and take the spectra around these emission lines with certain exposure and at chosen interval, generally referred as ``sit and stare mode''. The second mode of observations is to move the coronal image on the slits at certain chosen steps in multiple of 10 microns and at chosen time interval using linear scan mechanism (LSM) and record the spectral image at each step, referred as ``Raster scan''. The analysis procedure of spectral images is the same in both the cases. In case of ``sit and stare'' mode one investigates temporal variations at the locations of the slits, whereas in raster scan observations, one studies the spatial variation over the corona by making the 2-dimensional (2-D) images of solar corona using spectra. One can also determine the temporal variations by taking multiple raster scans. It takes longer time to make raster scan and thus one can study slow variations on 2-D image. In case of sit and stare mode one can study relatively faster variations but on limited coronal region along the slits. It is also planned to take the spectra in the sit and stare mode at longer interval ($\sim$1 minute) for longer periods to study the CMEs. The Gaussian fit to the emission line profile at each spatial location in the corona will be made to compute the peak (intensity), line-width (FWHM) and central position of the peak. From these data, one will be able to generate the intensity, velocity in the line-of-sight and line-width maps of solar corona including CME. The images in continuum and velocity maps will help to determine the true velocity of CME. We plan to make catalog of various features of CMEs and put in website for the scientific use of the data.

\subsection{ \textbf{Dark, flat-field and geometrical corrections }}

\begin{figure}[!htbp]
\begin{center}
\includegraphics[width=5.4in, height=3.5in,scale=0.5]{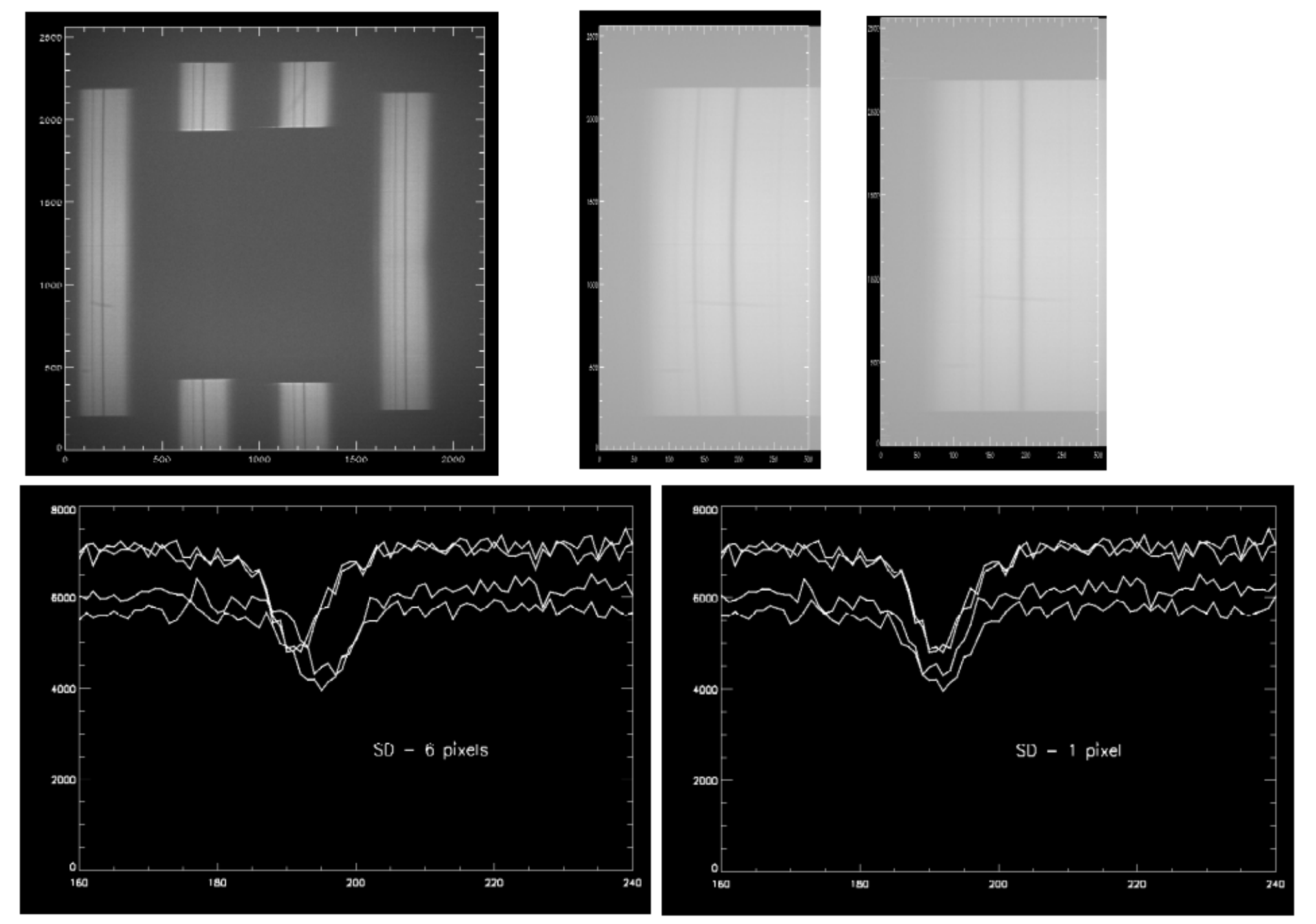}
\caption{The left panel of figure shows the spectrum of the solar disk taken with VELC at 5303\r{AA}. Middle panel in the top row shows the spectrum due to one slit (left) of the 4 slits. Right panel shows the spectrum after the geometrical corrections.  The left and right panels in the bottom row show the profiles of the absorption line before and after the geometrical corrections at various spatial locations. }
\label{fig:11}
\end{center}
\end{figure}

Using the dark, detector$'$s flat-field spectra and other information about the calibration of detectors such as hot and dead pixels, the spectra will be corrected (\citet{singh2022}). Sometimes the recorded spectrum shows curvature due to optics and tilt in the spectra because of mounting of detector in the instrument. The absorption lines in the solar disk spectrum show very small velocity ($<$ 1 km$/$sec) as compared to the velocity of plasma in dynamic solar corona. Further, disk spectrum due to scattered light  of the sun at different locations along the slit shows much less velocity because of contribution from the whole solar disk. Generally, the coronal spectrum shows two parts, one due to disk scattered light in the instrument and other because of emission from the hot coronal plasma. Using the absorption line in the spectrum, the spectra of various locations are shifted and aligned to a reference spectrum (chosen spectra at the center of slit) to correct for the curvature and tilt in all the spectra. The left panel of Figure \ref{fig:11} shows the spectrum of the solar disk taken with VELC at 5303\r{AA}. M2 mirror of the VELC was illuminated with sunlight using a fiber bundle to take the disk spectra. There are 4 spectra due to 4 slits of the spectrograph. The missing part of the spectra due to 2 middle slits is because of hole in the M2 mirror. To correct for the geometrical corrections spectrum due to extreme left slit was separated as seen in the middle panel of the top row of the figure. The enlarged view of the spectrum shows the curvature in the absorption line. The profile at a spatial location of 1200 row was chosen and all the spectra at other locations were shifted such that minimum of absorption line coincide with that of 1200 location. Right panel in the top row shows the spectrum after the geometrical corrections indicating that curvature in the spectrum has been corrected.  Further, the left panel in the bottom row shows that the minimum of the absorption line at different locations differ by 5 – 6 pixels but after the geometrical corrections the minimum of absorption line at various spatial locations coincide as seen in the right panel in the bottom row of the figure. Similarly, we make the geometrical corrections  for the spectra due to other slits choosing the respective reference location. 

\subsection{ \textbf{Conversion of pixel scale to wavelength scale }}

We have developed a code to create a file having the values of pixel versus wavelength by comparing the absorption lines in the disk or coronal spectrum with the atlas spectrum (https$:$//nispdata.nso.edu/ftp/pub/atlas/fluxatl/) of the sun. In this process two absorption lines are identified in the disk or coronal spectra and the corresponding lines are selected in the solar atlas spectrum. After comparing the average centers (away from active regions) of these absorption lines, the code computes the wavelength of each pixel and the dispersion of the spectrum.

\subsection{ \textbf{ Correction for Narrow-band filter transmission for Multi-slit observations}}

\begin{figure}[!htbp]
\begin{center}
\includegraphics[width=\textwidth]{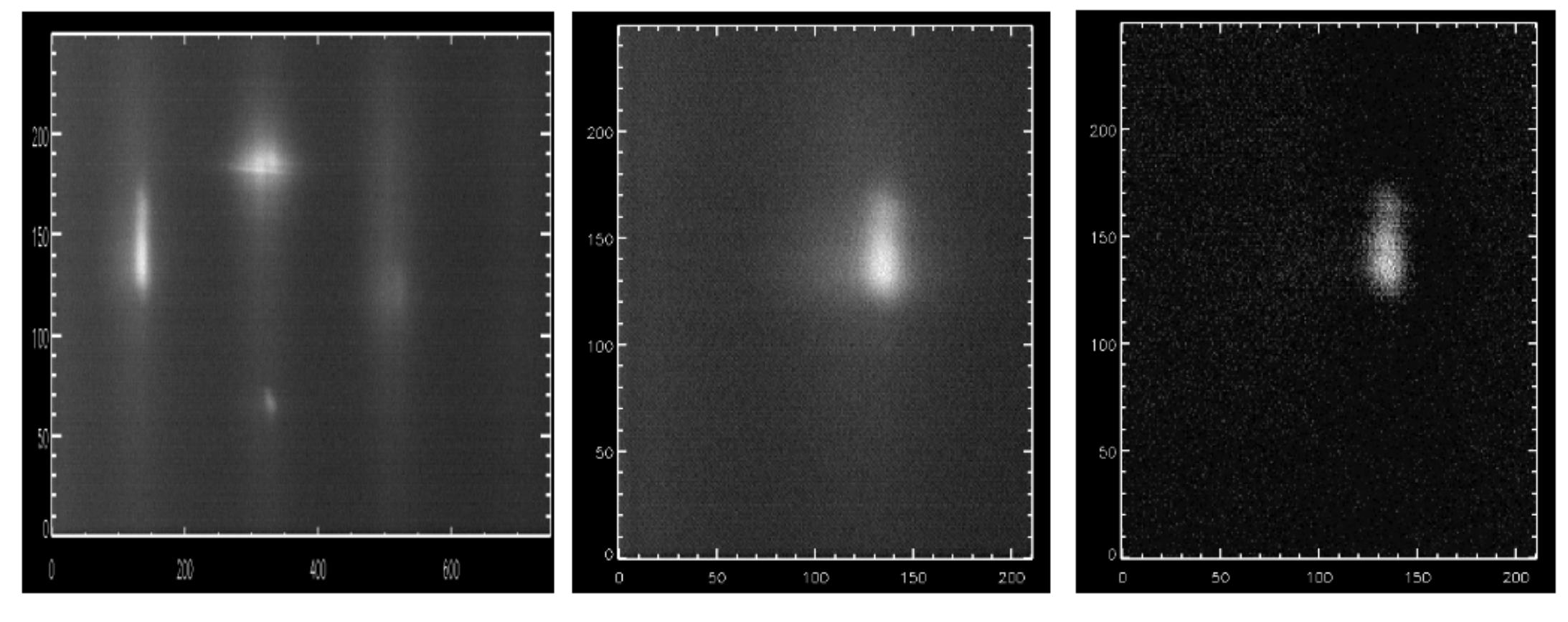}
\caption{Left panel of the figure shows the spectra due to 3 slits obtained during the total solar eclipse of July 11, 2010. The spectrum due to 4th slit at extreme right is very faint and does not show emission line. Middle and right panels of the figure show spectrum of the extreme left slit before and after compensating the transmission of the narrow band filter. }
\label{fig:12}
\end{center}
\end{figure}

The use of narrow band filters to avoid the overlap of spectra due to one slit with other slits complicates the data analysis. After the dark, flat-field and geometrical corrections as shown in figure \ref{fig:12}, the spectra need to be compensated for the transmission curve of the filter. It is easy to handle the analysis by separating the spectra due to each slit and later combining the results. Here, we have considered the multi-slit spectra obtained in 530.3 nm [Fe xiv] emission line during the total solar eclipse of 2010 at Easter Island (\citet{samanta2016}). The exponential decrease in coronal intensity with increasing solar radii adds to complications. After determining the contribution due to transmission filter at each location in solar corona and at each wavelength for continuum part of the spectra, the spectra were corrected for the transmission profile of the filter to make uniform background at continuum part. Left panel of Figure \ref{fig:12} shows the spectra due to 3 slits obtained during the total solar eclipse of July 11, 2010. Middle and right panels of the figure show spectrum of the extreme left slit before and after compensating the transmission of the narrow band filter.

\subsection{ \textbf{Scattered light correction}}

The coronal spectra include the disk light spectra due to scattering of solar disk light by earth's atmosphere in case of ground base observations or by instrument while observing from space . In some cases, absorption line of the disk spectrum is blended with the emission line. For example, the 789.19 nm absorption line lies at the centre of [Fe xi] emission line at 789.2 nm and another absorption line at 530.27 nm at blue wings of [Fe xiv] emission line. The contribution of the continuum and these absorption lines to the emission line needs to be removed to determine profile of the emission line.

\begin{figure}[!htbp]
\begin{center}
\includegraphics[width=\textwidth]{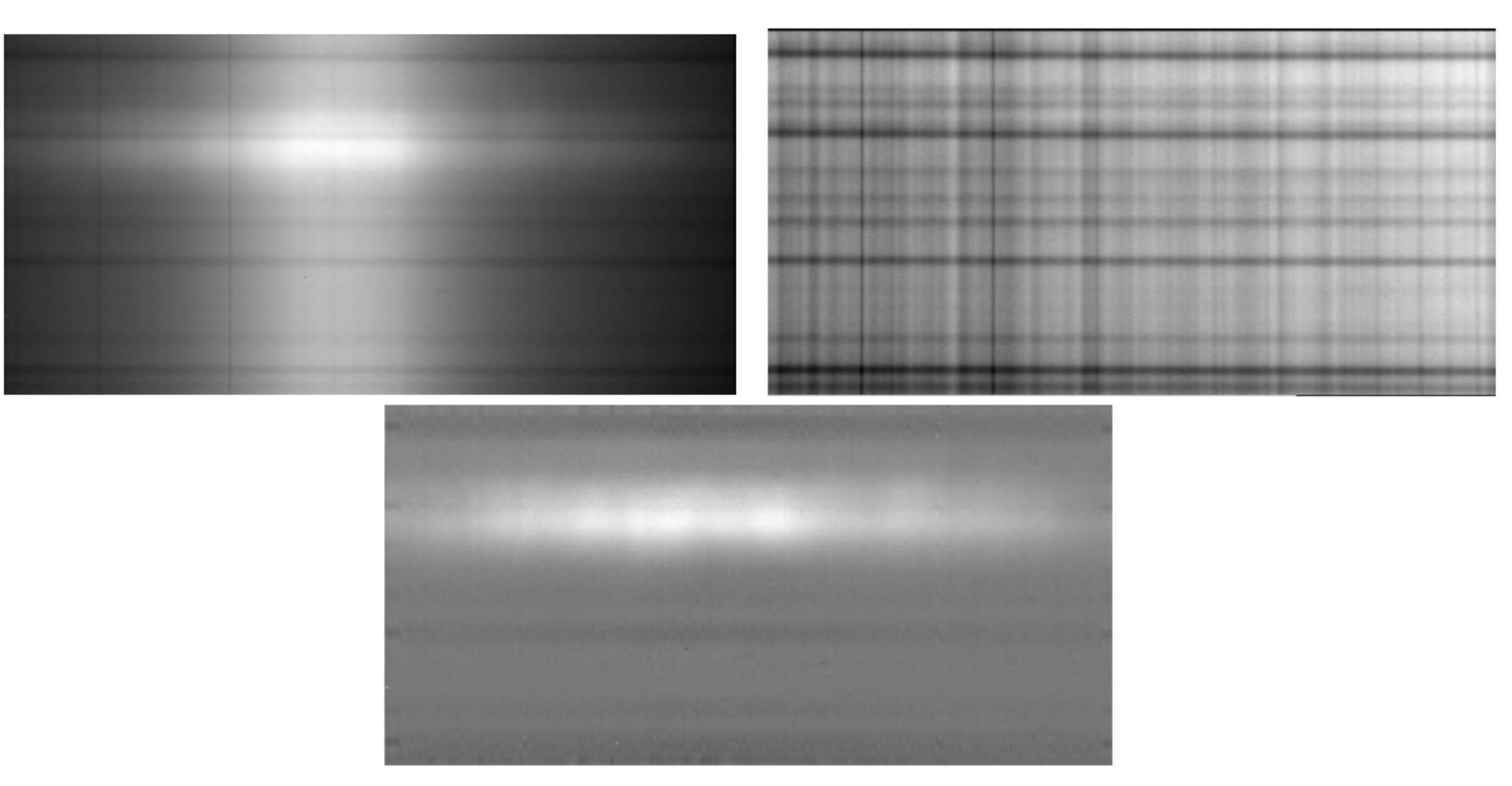}
\caption{Top-left panel of the figure shows the coronal spectra before scattering correction, top-right panel shows the disk spectra and bottom panel shows the spectra after removing the scattered light component due to sky and instrument. The 7892 \r{AA} [Fe xi] emission line becomes dominant with very faint absorption line spectrum in the background. These observations were obtained with 25-cm coronagraph at Norikura observatory.}
\label{fig:13}
\end{center}
\end{figure}

Left and right side of top panel in the figure \ref{fig:13} show the coronal and disk spectra around 7892\r{AA} [Fe xi] emission line, respectively, obtained with 25-cm coronagraph at Norikura observatory. Using the disk spectra and absorption lines we remove the contribution of scattered sunlight and resulting emission line spectra is seen in the bottom panel of the figure \ref{fig:13}. This panel clearly shows the emission line at 7892\r{AA}. There is still a signature of absorption lines against the uniform background. The intensity of remanant of absorption lines is very less and do not have any effect in fitting the Gaussian profile to the emission line and in the determination of emission line parameters, such as peak intensity, central wavelength and line-width. The developed code will be applied to the spectroscopic observation obtained with VELC.

\subsection{ \textbf{Determination of emission line parameters}}

\begin{figure}[!htbp]
\begin{center}
\includegraphics[width=5.4in, height=5.4in,scale=0.5]{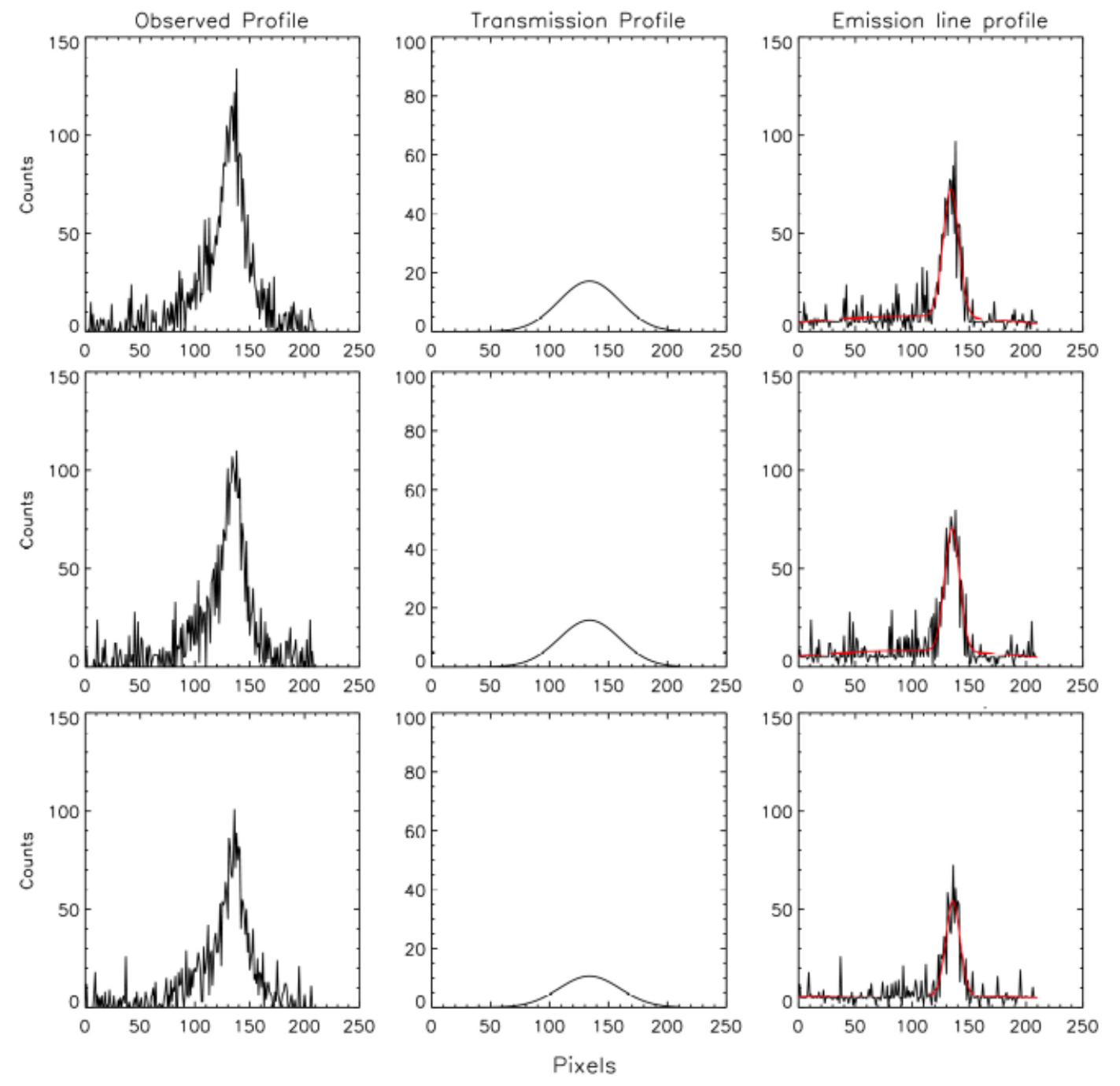}
\caption{The figure shows the observed profiles at three representative spatial locations obtained during the total solar eclipse of July 11, 2010 using multi-slit spectrograph. The middle and right side columns show the contribution of transmission profile of filter and the remnant emission line profile after compensating the transmission profile of the narrow band filter. A Gaussian fit to the emission line profile is  shown in red color.}
\label{fig:14}
\end{center}
\end{figure}

\begin{figure}[!htbp]
\begin{center}
\includegraphics[width=4.4in, height=3.2in,scale=0.5]{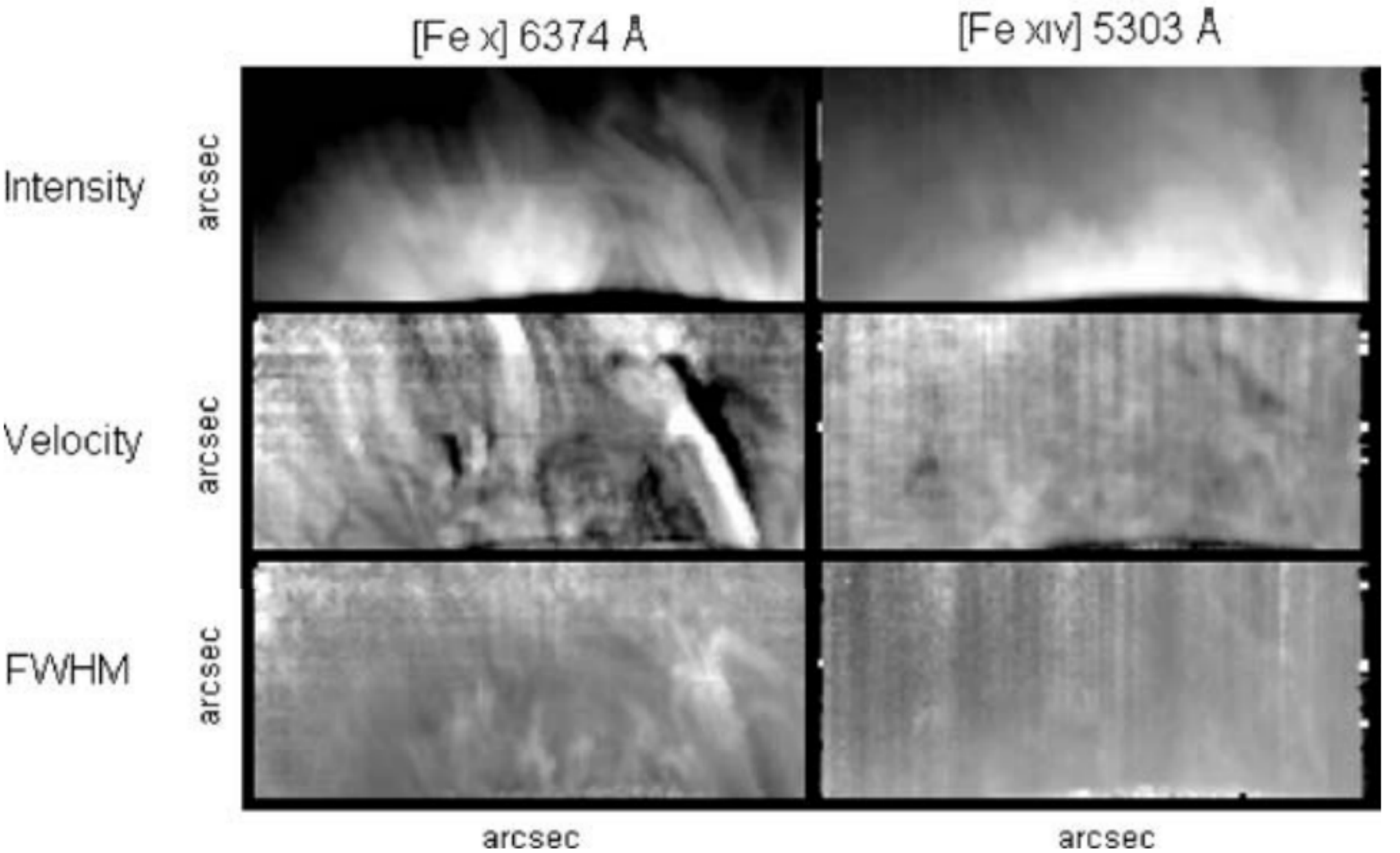}
\caption{The figure shows the generated intensity, doppler velocity and FWHM images of the scanned region. Top panels show the intensity distribution of the 6374\r{AA} (left) and 5303\r{AA} (right) coronal emission lines observed simultaneously in a coronal region of 240 × 500 size on 25 October 2003 (22:09 UT). Middle and bottom panels show the Doppler velocity and FWHM of the emission lines, respectively. }
\label{fig:15}
\end{center}
\end{figure}

The faint absorption lines (Figure \ref{fig:13}) seen in the emission line spectra are due to small residuals and do not have any impact on the determination parameters of emission line using Gaussian fit to the observed profiles. Defining the approximate centre of emission line, interval of the Gaussian fit, information about the pixel to wavelength scale conversion, we determine the peak intensity, full width half   maximum (FWHM) and Doppler velocity at all the spatial locations along the slit. First column of the Figure \ref{fig:14} shows the observed profiles at three representative spatial locations obtained during the total solar eclipse of July 11, 2010 using multi-slit spectrograph (\citet{samanta2016}). The middle and right side columns show the contribution of transmission profile of filter and the remnant emission line profile after compensating the transmission profile of the narrow band filter. A Gaussian fit to the emission line profile is also shown to compute the line-width, intensity and line-of-sight velocity at that spatial location. The values of peak intensity, position of peak intensity, FWHM in pixels and FWHM in Angstrom of the emission line after the correction for the instrumental profile are given in Table \ref{tab:1}. The spectrum appears noisy because of very short exposure time to study the temporal oscillations. This analysis will be done at each spatial location along the slits. It may be noted the code has a provision, not to consider certain number of pixels while making a Gaussian fit to the emission line because of residual signal at those pixels due to absorption line. In case of raster scan observations these parameters will be combined to make intensity, Doppler and line-width maps of the scanned region. The data of all the 4 slits will be combined to generate image of the observed solar corona. The Figure  \ref{fig:15} shows an example of a coronal region observed with 25-cm coronagraph in a similar procedure.

\begin{table}[ht!]
\begin{center}
\caption{The values of  spatial location in the corona, peak intensity, position of peak in intensity, FWHM of emission line and FWHM in \r{AA} after correcting for the instrumental profile are listed.}
\label{tab:1}
\scalebox{0.8}{%
\begin{tabular}{ |c|c|c|c|c| }
 \hline
 
\textbf{Sp.Location} & \textbf{Peak Intensity} & \textbf{Peak Position} & \textbf{FWHM from Gaussian fit} & \textbf{FWHM(\r{AA}) Instrument corrected} \\
 \hline
Row-135 & 65.21 & 133.92 & 16.29 & 0.98 \\
\hline
Row-145 & 62.69 & 134.92 & 15.5 & 0.93 \\
\hline
Row-155 & 48.04 & 135.65 & 14.57 & 0.87 \\
\hline
\end{tabular}}
\end{center}

\end{table}

\subsection{ \textbf{Alignment and re-scaling of images }}

The CMOS detectors to record Fe[xi] and Fe[xiv] lines and IR detector for Fe[xiii] emission line being used have different pixel spatial scales, 1.25 arcsec $/$ pixel for CMOS and 4.8 arcsec $/$ pixel for IR detector.  The images made from the observed spectra need to be aligned with each other and made of equal format, that means of equal spatial scale by adjusting their spatial scales to compare the parameters of different emission lines. This will be achieved by taking the spectra with a cross wire put on the slits of spectrograph. A code has been developed and tested to do it.

\subsection{ \textbf{ Temperature and Doppler maps of solar corona}}

The intensity of these coronal lines is temperature sensitive as the abundance of respective ions depends on the temperature of the plasma. By taking the ratio of intensity of these lines e.g., Fe[xiii] $/$ Fe[xi] and Fe[xiv] $/$ Fe[xi] we shall generate the temperature maps of the solar corona. Using this temperature map and line width information, we shall be able derive the non-thermal component of the plasma at each location of the solar corona and thus generate Doppler map of solar corona. Multi-temperature intensity, velocity, line-widths, and Doppler maps are expected to provide detailed physical and dynamical nature of solar corona and coronal loops.

\subsection{ \textbf{Alignment of maps }}

The location of a point in the solar corona is generally defined in terms of solar radii from the centre of sun and the angle measured from the north pole of sun towards east. Hence, images will be rotated considering the ``Yaw'' angle of the satellite and ``P'' angle of the sun at the time of observations to make the north pole of the sun vertical and east on the left side of the image as is the general norm. The image will also be moved depending on the ``roll and pitch'' angles of the satellite at the time of observations.

\section{Analysis of spectro-polarimetric data}

While making the spectro-polarimetric observations of the solar coronal in Fe[xiii] emission line, other two channels may also be recording the spectra in Fe[xi] and Fe [xiv] emission lines. There will be two spectra in Fe[xiii] line for each slit because of polarizing beam splitter. The analysis of the spectra till the generation of emission line profiles will be done in a similar way as partly explained in \citet{nagaraju2021} and \citet{sasikumar2022}. To derive the I, Q, U and V parameters, a complete methodology will be adopted which will be described separately.

\section{Availability of software codes }

It is planned that information about codes will be shared with data users before and after the launch of the payload by arranging meetings at Indian Institute of Astrophysics, Bengaluru. Some training will also be given to the participants. These codes need to be verified with the actual data and working of the instrument during the PV (payload verification) phase. After making the required changes in codes and confirming the proper working of codes, these will be put in public domain. Again, information and training to the users will be provided in workshops during the PV and GT phase. It may be noted that proposal submission form to make the observational plan, is being worked and tested. It has gone through number of revisions because of changes in hardware and control electronics. The details of proposal submission form will be shared in all the proposed workshops and meetings.

\section{Acknowledgements }
We thank all the Scientists$/$Engineers at the various centres of ISRO such as URSC, LEOS, SAC, VSSC etc. and Indian Institute of Astrophysics who have made great contributions to the mission to reach at the present state. We gratefully acknowledge the financial support from ISRO for this project. The coronal spectroscopic  data used here was obtained by Prof. Jagdev Singh at Norikura observatory, Japan. The SOHO$/$LASCO data used here are produced by a consortium of the Naval Research Laboratory (USA), Max-Planck-Institut for Sonnensystemforschung (MPS, Germany), Laboratoire d$'$Astronomie (LAS, France), and the University of Birmingham (UK). SOHO is a project of international cooperation between ESA and NASA.


\begin{thebibliography}{}

\bibitem[\protect\citeauthoryear{Brueckner et~al (1995) }{1995}]{brueckner1995}
Brueckner, ~G. E., Howard, ~R. A., Koomen, ~M. J.  1995, The Large Angle Spectroscopic Coronagraph (LASCO). \textit{Solar Physics}, \textbf{162},  357.
\\

\bibitem[\protect\citeauthoryear{Ichimoto et~al (1999)} {1999}]{ichimoto1999}
Ichimoto, ~ Kiyoshi., Noguchi, ~Motokazu.,  Tanaka, ~Nobuyuki., Kumagai, ~Kazuyoshi., Shinoda, ~Kazuya., Nishino, ~Tetsuo., Fukuda, ~Takeo., Sakurai, ~Takashi., Takeyama, ~Norihide. 1999, A New Imaging System of the Corona at Norikura, \textit{PASJ}, \textbf{51}, 383 - 391.
\\

\bibitem[\protect\citeauthoryear{Kumar et~al (2018)} {2018}]{kumar2018}
 Kumar, ~N.,  Raghavendra Prasad, Budihal.,  Singh, Jagdev.,  Venkata, ~S. N. 2018, Optical design of visible emission line coronagraph on Indian space solar mission Aditya-L1. \textit{Experimental Astronomy}, \textbf{45}.
\\

\bibitem[\protect\citeauthoryear{Nagaraju et~al (2021) }{2021}]{nagaraju2021}
Nagaraju, ~K., Prasad, ~B.R., Hegde, ~B.S., et al. 2021, Spectropolarimeter on 	board the aditya-L1, polarization modulation and demodulation. \textit{Applied optics}, \textbf{60}, 8145-8153.
\\

\bibitem[\protect\citeauthoryear{Prasad et~al (2017)} {2017}]{prasad2017}
 Raghavendra Prasad, Budihal.,  Banerjee, Dipankar.,  Singh, Jagdev.,  Subramanya, Nagabhushana.,  Kumar, Amit.,  Kamath, ~P.,  Kathiravan, ~S.,  Venkata, ~S.N,  Rajkumar, ~N.,  Venkatasubramanian, Natarajan ., Juneja, Madhur., Somu, Pawan.,  Pant, Vaibhav., Shaji, Nigar., Sankarsubramanian, ~K.,  Patra, Asit.,  Venkateswaran, R.,  Adoni, Abhijit.,  Narendra, ~S.,  Jaiswal, Bhavesh. 2017, Visible Emission Line Coronagraph on Aditya-L1, \textit{Current Science}, \textbf{113}, 613-615.
\\

\bibitem[\protect\citeauthoryear{Samanta et~al (2016) }{2016}]{samanta2016}
Samanta, ~T., Singh, ~J., Sindhuja, ~G.,  Banerjee, ~D. 2016, Detection of High-Frequency Oscillations and Damping from Multi-slit Spectroscopic Observations of the Corona, \textit{Solar Physics}, \textbf{291}, 155.
\\

\bibitem[\protect\citeauthoryear{Sasikumar et~al (2022) }{2022}]{sasikumar2022}
Sasikumar Raja, ~K., Venkata, ~S.N., Singh, ~J., Prasad, ~B.R. 2022, Solar coronal magnetic ﬁelds and sensitivity requirements for spectropolarimetry channel of VELC onboard Aditya-L1. \textit{Advances in Space Research}, \textbf{69}, 814-822.


\bibitem[\protect\citeauthoryear{Singh et~al (2004)} {2004}]{singh2004}
Singh, ~J., Takashi Sakurai., Kiyoshi Ichimoto., Tetsuya Watanabe. 2004, Complex variations in the line-intensity ratio of coronal emission lines with height above the limb, \textit{The Astrophysical Journal}, \textbf{617}, L81–L84.
\\

\bibitem[\protect\citeauthoryear{Singh et~al (2006)} {2006}]{singh2006}
Singh, ~J., Takashi sakurai.,  Kiyoshi Ichimoto. 2006, Do the line widths of coronal emission lines increase with height above the limb?. \textit{The Astrophysical Journal}, \textbf{639}, 475–483.
\\

\bibitem[\protect\citeauthoryear{Singh et~al (2011)} {2011}]{singh2011}
Singh, Jagdev., Raghavendra Prasad, Budihal.,  Venkatakrishnan, ~P.,  Sankarasubramanian, ~K.,  Banerjee, Dipankar.,  Bayanna, ~A.,  Mathew, Shibu.,  Murthy, Jayant.,  Subramaniam, Prasad.,  Rajaram, Ramesh.,  Kathiravan, ~S.,  Subramanya ., Nagabhushana, ~K., Mahesh., Manoharan, ~P.,  Uddin, Wahab.,  Sripadmanaban, Sriram.,  Kumar, Amit ., Srivastava, ~N.,  Rao, Koteswara.,  Patra, Asit. 2011, Proposed visible emission line space solar coronagraph, \textit{Current Science}, \textbf{100}.
\\

\bibitem[\protect\citeauthoryear{Singh et~al (2011)} {2011}]{singh2011a}
Singh, ~J., Hasan, ~S.S., Gupta, ~G. R., Nagaraju, ~K.,  Banerjee, ~D. 2011, Spectroscopic Observation of Oscillations in the Corona During the Total Solar Eclipse of 22 July 2009, \textit{Solar Phys}, \textbf{270}, 213–233.
\\

\bibitem[\protect\citeauthoryear{Singh et~al (2019) }{2019}]{singh2019}
Singh, ~J., Raghavendra Prasad, ~B., Venkata, ~S., Kumar, ~A. 2019, Exploring the outer emission corona spectroscopically by using Visible Emission Line Coronagraph (VELC) on board ADITYA-L1 mission, \textit{Advances in Space Research}, \textbf{64}, 7, 1455–1464.
\\

\bibitem[\protect\citeauthoryear{Singh et~al (2022) }{2022}]{singh2022}
Singh, ~J., Raghavendra Prasad, ~B., Chavali Sumana, Amit Kumar, Varun Kumar, Muthu Priyal, Venkata, ~S.N. 2022, Data pipeline architecture and development for VELC onboard Space Solar Mission AdityaL1, \textit{Advances in Space Research}, \textbf{69}, 2601–2610.
\\

\bibitem[\protect\citeauthoryear{Venkata et~al (2017) }{2017}]{venkata2017}
Venkata, ~S.N., Prasad, ~B.R., Nalla, ~R.K., Singh, ~J. 2017, Scatter studies for visible emission line coronagraph on board ADITYA-L1 mission. \textit{J. Astron. Telescopes Instrum. Syst.}, \textbf{626},  3.
\\

\bibitem[\protect\citeauthoryear{Venkata et~al (2021) }{2021}]{venkata2021}
Venkata, ~S.N., Prasad, ~B.R., Singh, ~J. 2021, Spectropolarimetry Package for Visible Emission Line Coronagraph (VELC) on board Aditya-L1 Mission. \textit{Experimental Astronomy}, \textbf{53}, 71-82.
\\
	
\bibitem[\protect\citeauthoryear{ Wülser et~al (2018) }{2018}]{wulser2018}
Wülser, ~JP., Jaeggli, ~B., De Pontieu., et al.  2018, Instrument Calibration of the Interface Region Imaging Spectrograph (IRIS) Mission. \textit{Solar Physics}, \textbf{293}, 149.
\\
 
\end{thebibliography}
\end{document}